\documentclass[pra,twocolumn]{revtex4-1}
\usepackage{hyperref} 

\usepackage{graphicx} 
\usepackage{subfig}
\usepackage{color}
\usepackage{filecontents}
\usepackage{soul}
\usepackage{bbold}

\newcommand{\be}{\begin{equation}}
\newcommand{\ee}{\end{equation}}
\newcommand{\bea}{\begin{eqnarray}}
\newcommand{\eea}{\end{eqnarray}}
\newcommand*{\myeqref}[2][Eq.~]{%
  \hyperref[{#2}]{#1(\ref*{#2})}%
}
\def\equationautorefname#1#2\null{%
  Eq.#1(#2\null)%
}

\usepackage{dcolumn}
\usepackage{bm}
\usepackage{amssymb,amsmath}
\usepackage{color}
\usepackage{float}
\usepackage{tikz}
\usetikzlibrary{arrows}

\definecolor{DarkGreen}{rgb}{0,0.6,0.2}

\begin{document}
\title{On the optical nonreciprocity and slow light propagation in coupled spinning optomechanical resonators}

\author{Imran M. Mirza$^1$, Wenchao Ge$^{2}$, and Hui Jing$^{3}$}
\affiliation{$^{1}$Macklin Quantum Information Sciences, 
Department of Physics, Miami University, Oxford Ohio 45056, USA\\
$^{2}$Institute for Quantum Science and Engineering,  Department of Physics and Astronomy, Texas A\&M University, College Station, TX 77843, USA\\
$^{3}$Key Laboratory of Low-Dimensional Quantum Structures and Quantum Control of Ministry of Education, Department of Physics and Synergetic Innovation Center for Quantum Effects and Applications,
Hunan Normal University, Changsha 410081, China}
\email{imranmajidmirza@gmail.com}

\begin{abstract}
We study the optical transmission characteristics of pump-probe driven spinning optomechanical ring resonators coupled in a series configuration. After performing the steady-state analysis valid for an arbitrary number of resonators, as an example, we discuss the two-resonator problem in detail. Therein, we focus on how changing the optical Sagnac effect due to same or opposite spinning directions of resonators can lead to enhanced, non-reciprocal and delayed probe light transmission. This work can help in devising spin degree of freedom based novel devices of manipulating light propagation in quantum networks and quantum communication technologies.
\end{abstract}

\maketitle

\section{Introduction}
Nonreciprocal optical devices have gained a lot of attraction due to their ability to break the symmetry of an experiment under the source and detector exchange even in the presence of noise \cite{deak2012reciprocity, feng2011nonreciprocal, kamal2011noiseless}.  Prime examples of such devices are optical circulators \cite{scheucher2016quantum} and isolators \cite{sayrin2015nanophotonic} which have found applications in quantum networking, one-way optical communication protocols and topological photonics \cite{tian2017nonreciprocal, lodahl2017chiral, metelmann2017nonreciprocal}. For a recent experiment on optomechanical circulators see \cite{shen2018reconfigurable}. Some of the underlying physical effects utilized in these devices include optical nonlinearity \cite{liu2014regularization}, magneto-optical crystal based Faraday rotation \cite{chin2013nonreciprocal, floess2017plasmonic}, reservoir engineering \cite{metelmann2015nonreciprocal} and photonic Aharonov-Bohm effect \cite{yuan2015achieving}.

In addition to nonreciprocity, tunable slow and fast light propagation \cite{boyd2009slow, pant2012photonic} is another useful mechanism to provide on-demand photon transmission in chip devices. Typically, in quantum optical settings, the slow and fast light is achieved through the phenomenon of electromagnetically induced transparency (EIT) \cite{fleischhauer2005electromagnetically} in which a strong pump field driven three-level atomic medium becomes transparent to a weak probe field due to quantum coherence. More recently EIT has been studied in disparate setups, for instance in metamaterials \cite{liu2009plasmonic}, circuit quantum electrodynamics \cite{abdumalikov2010electromagnetically}, waveguide quantum electrodynamics \cite{witthaut2010photon, mirza2018influence} and optomechanics \cite{safavi2011elect}. Analogous to EIT in other physical setups, optomechanics also manifests EIT due to interference among different decay channels incorporating mechanical side-bands. This phenomenon in optomechanics is commonly referred to optomechanically induced transparency (OMIT) \cite{weis2010optomechanically, kronwald2013optomechanically, agarwal2010OMIT}.

In this context, cavity quantum optomechanics \cite{aspelmeyer2014cavity} is particularly a noticeable example due to its ability to host both nonreciprocity \cite{manipatruni2009optical} and slow and fast light propagation \cite{jiang2013electromagnetically} in the same setup. More recently, hybrid atom-optomechanical systems (with single atoms, Bose-Einstein condensates and Kerr-type nonlinear medium coupled to optomechanical systems) \cite{rogers2014hybrid, jiao2018optomechanical, mirza2015real, mirza2016strong} have made these studies more interesting due to coherent coupling among different degrees of freedom. The presence of atom-like system can lead to for instance improved ground state cooling of a mechanical oscillator \cite{zeng2017ground}, steady-state mechanical squeezing \cite{wang2016steady}, appearance of novel correlations \cite{restrepo2017fully} and qubit-assisted enhancement of optomechanical interaction \cite{pirkkalainen2015cavity}.

However, besides coupling emitters to an optomechanical cavity, there are other fascinating possibilities of introducing hybrid degrees of freedom in quantum optomechanics \cite{koutserimpas2018nonreciprocal,el2013chip}. In particular, recently L{\"u} et al. have investigated a fiber coupled cavity optomechanical system where the cavity (ring-resonator in their study) is capable of spinning/rotating as well \cite{lu2017optomechanically}. They have shown that spin degrees of freedom can considerably modify the probe light transmission and even can introduce nonreciprocity with slow and fast light propagation. In this same regard, Maayani et al.  have demonstrated that optical non-reciprocal transmission can be achieved by coupling spinning resonators with flying couplers \cite{maayani2018flying}. For further very recent developments (such as nanoparticle sensing and nonreciprocal photon blockade) with setups involving spinning resoantors we direct reader to the references \cite{huang2018nonreciprocal,*hui2018Nano}.

Motivated by L{\"u} and Maayani et. al studies, in this work we consider multiple spinning resonators that are coupled in a series configuration. After presenting the analytic work valid for an arbitrary number of resonators, we concentrate on the problem of two-resonators where we basically ask the question that how the presence of different spinning directions can impact our ability to alter the probe light transmission. We find that even in the case of two-resonators, different spin directions can considerably change the transport of probe photons. For instance, for a set of experimentally feasible parameters, we notice by selecting clockwise spinning of both resonators one can observe the emergence of a ``W" like transmission pattern for negative detunings and an enhancement in the transmission by a factor of 3 as compared to the single spinning resonator case at spinning rates of $100$kHz and detuning of $10$MHz. Moreover, we also point out that by choosing same spinning directions in both resonators one can achieve both higher transmission as well as slow light propagation without requiring any optical non-linearity and magneto-optical effect.

The paper is organized as follows. In section II we outline the setup description. Followed by this, in Section III we model the system and present the steady-state analysis of probe transmission for an arbitrary number of resonators. Section IV is devoted to the discussion of results for a two-resonator problem. Finally, in Section V, we close with a summary of our results and a glimpse of future research directions.

\section{Setup and Experimental Feasibility}
We consider a vertical array of $N$ number of series-coupled optomechanical spinning microresonators as depicted in Fig.~1. The bottom resonator is coupled to an optical (tapered) fiber which guides two incoming fields namely a weak probe field with amplitude $\varepsilon_{p_{1}}$ and a strong pump field with amplitude $\varepsilon_{l_{1}}$. The respective powers of incoming probe and pump fields $P_{in}$ and $P_{l_{1}}$ are related to these amplitudes through $\varepsilon_{p_{1}}=\sqrt{\frac{P_{in}}{\hbar\omega_{p}}}$ and $\varepsilon_{l_{1}}=\sqrt{\frac{P_{l_{1}}}{\hbar\omega_{l_{1}}}},$ where $\omega_{p}$ and $\omega_{l_{1}}$ are the frequencies of the probe and pump light, respectively. In our model any $j$th resonator ($1\leq j\leq N$) in the configuration is driven by a strong pump/control field with frequency $\omega_{l_{j}}$ which also excites a breathing mechanical-mode with frequency $\omega_{m_{j}}$ and effective mass $m_{j}$. We also assume each resonator supports a single isolated optical mode with frequency $\omega_{c_{j}}$ in the $j$th resonator. This optical mode is then coupled with the mechanical mode (in the same resonator) through the standard non-linear optomechanical interaction \cite{aspelmeyer2014cavity} (with strength parameter $\xi_{j}$). The series coupled configuration of spinning optomechanical resonators considered in this work is similar to the setup mentioned in \cite{li2010coupled} where two coupled optical modes with asymmetric waveguide interaction was studied.

\begin{figure}
   \includegraphics[width=3.5in, height=6.85in]{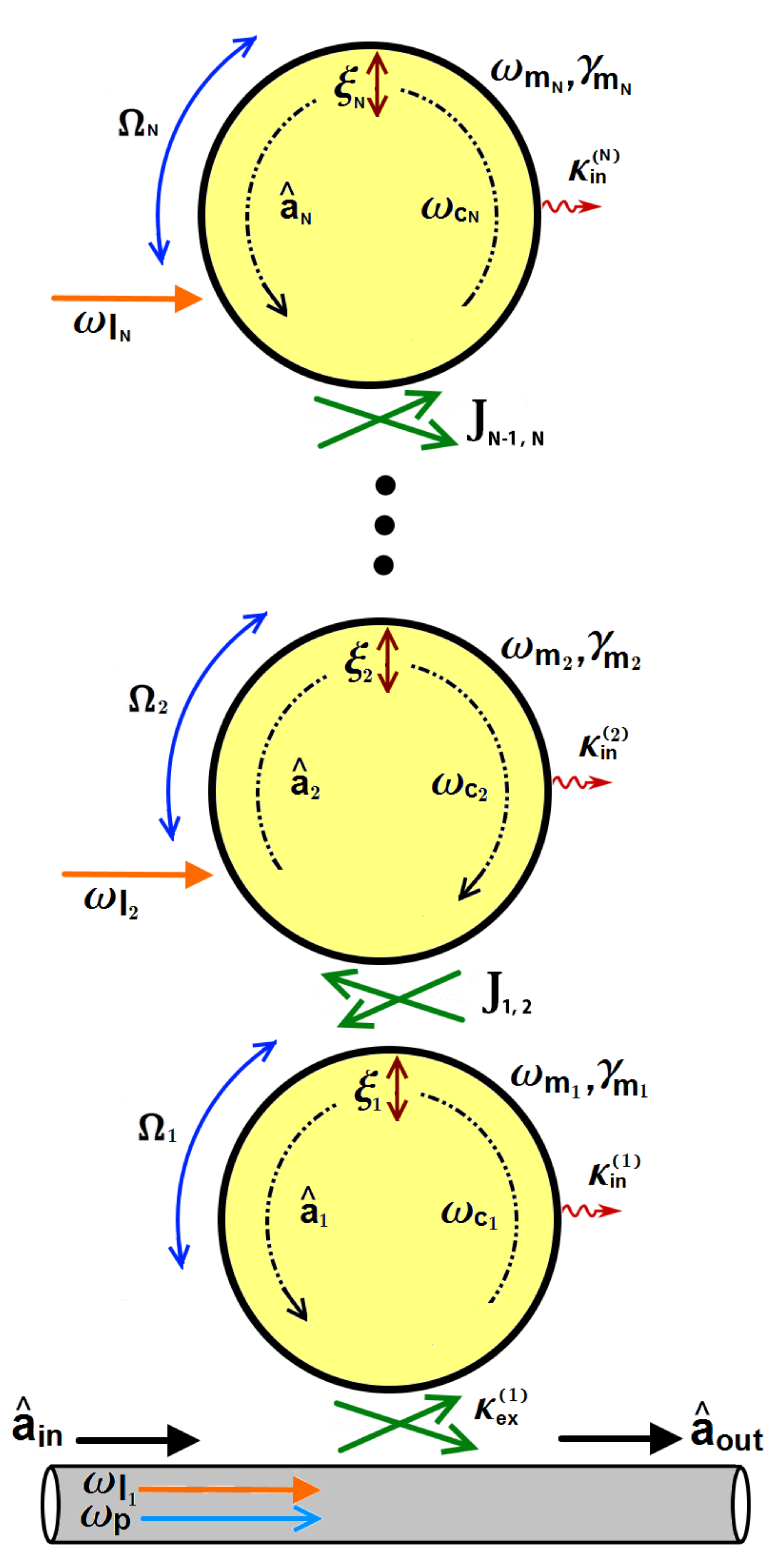}  
\captionsetup{
  format=plain,
  margin=1em,
  justification=raggedright,
  singlelinecheck=false
}
 \caption{(Color online) A chain of series-coupled spinning optomechanical ring resonators.}\label{Fig1}
\end{figure}

To make our system more realistic, we incorporate the internal cavity loss (represented by $\kappa^{(j)}_{in}$). For the bottom resonator, in addition to the internal loss we also consider external loss (given by $\kappa^{(1)}_{ex}$ or simply $\kappa_{ex}$) which describes the evanescent coupling of the resonator with the fiber. Moreover, we consider all resonators to be capable of spinning with the spinning/rotating frequency of the $j$th resonator represented by $\Omega_{j}$. Conventionally we take $\Omega_{j}>0$ for clockwise rotary direction. In this work, we consider rotation speeds up to $100$kHz which seems challenging from the point of view of keeping motional stability as well as efficient optical coupling. However, this is not very far from implementation if we take into account the continued improvement in the optical resonator technology. For instance, in a very recent study \cite{maayani2018flying} for a $1.1 $mm radiuas resonator a spinning speed of $2\pi\times 3,000$ rad.s$^{-1}$ has been achieved to perform non-reciprocal light transmission.\\
It is well-known that due to rotation, optical mode frequency suffers {\it Sagnac-Fizeau} shift \cite{malykin2000sagnac} which transforms
\begin{equation}
\begin{split}
&\omega_{c_{j}}\longrightarrow\omega_{c_{j}}+\Delta_{sag_{j}},\\
&\text{where}\hspace{2mm}\Delta_{sag_{j}}:=\frac{n_{j}r_{j}\Omega_{j}\omega_{c_{j}}}{c}\Bigg(1-\frac{1}{n_{j}}-\frac{\lambda_{j}}{n_{j}}\frac{dn_{j}}{d\lambda_{j}}\Bigg).
\end{split}
\end{equation}
$\Delta_{sag_{j}}$, $n_{j}$, $r_{j}$ are the Sagnac-Fizeau shift, refractive index and radius of the $j$the resonator. $c$ is the speed of light and $dn_{j}/d\lambda_{j}$ term represents small relativistic (dispersion) correction in the Sagnac-Fizeau shift. 

\section{Output probe light: theoretical description}
\subsection{Hamiltonian}
Transforming into a frame rotating with pump frequency $\omega_{l_{j}}$, the Hamiltonian (with $\hbar=1$) under the rotating wave approximation is expressed as
\begin{equation}
\label{eq:Ham}
\begin{split}
&\hat{H}=\hat{H}_{0}+\hat{H}_{int}+\hat{H}_{dr},\\
&\hat{H}_{0}=\sum^{N}_{j=1}(\Delta_{c_j}\hat{a}^{\dagger}_{j}\hat{a}_{j}+\frac{\hat{p}^{2}_{j}}{2m_{j}}+\frac{1}{2}m_{j}\omega^{2}_{m_{j}}\hat{x}^{2}_{j}+\frac{\hat{p}^{2}_{j\theta}}{2m_{j}r^{2}_{j}}),\\
&\hat{H}_{int}=-\sum^{N}_{j=1}\xi_{j}\hat{x}_{j}\hat{a}^{\dagger}_{j}\hat{a}_{j}+\sum^{N-1}_{j=1}J_{j,j+1}(a^{\dagger}_{j}\hat{a}_{j+1}+\hat{a}^{\dagger}_{j+1}\hat{a}_{j}),\\
&\hat{H}_{dr}=\sum^{N}_{j=1}i\sqrt{\kappa_{ex}}(\varepsilon_{l_{j}}\hat{a}^{\dagger}_{j}+\varepsilon_{p_{1}}\hat{a}^{\dagger}_{1}e^{-i(\omega_{p}-\omega_{l_{j}})t}-h.c.).
\end{split}
\end{equation}
Here $\Delta_{c_j}=\omega_{c_j}-\omega_{l_{j}}$ and we have assumed only $\hat{a}_{1}$-mode is driven with the probe field.  Non-vanishing commutation relations obeyed by different operators are given by
\begin{equation*}
[\hat{x}_{j},\hat{p}_{x_{k}}]=i\delta_{jk},[\hat{\theta}_{j},\hat{p}_{k\theta}]=i\delta_{jk}\hspace{2mm}\text{and}\hspace{2mm} [\hat{a}_{j},\hat{a}^{\dagger}_{k}]=\delta_{jk}.
\end{equation*}

\subsection{Heisenberg-Langevin equations of motion}
Relevant operators' equations of motion in the Heisenberg picture can be worked out using Eq.~(\ref{eq:Ham}). We obtain
\begin{equation}
\label{eq:HeisEq}
\begin{split}
&\frac{d\hat{a}_{j}(t)}{dt}=-i(\Delta_{c_{j}}-\xi_{j}\hat{x}_{j}-i\beta_{j})\hat{a}_{j}+\sqrt{\kappa_{ex}}(\varepsilon_{l_{j}}+\\
&\varepsilon_{p_{1}}e^{-i\eta_{j}t}\delta_{j1})-iJ_{j,j-1}\hat{a}_{j-1}-iJ_{j,j+1}\hat{a}_{j+1},\\
&\frac{d^{2}\hat{x}_{j}(t)}{dt^{2}}=\frac{\xi_{j}}{m_{j}}\hat{a}^{\dagger}_{j}\hat{a}_{j}-\omega^{2}_{m_{j}}\hat{x}_{j}+\frac{\hat{p}^{2}_{j\theta}}{m^{2}_{j}r^{3}_{j}}-\gamma_{m_{j}}\frac{d\hat{x}_{j}}{dt},\\
&\frac{d\hat{\theta}_{j}(t)}{dt}=\frac{\hat{p}_{j\theta}}{m_{j}r^{2}_{j}},\\
&\frac{d\hat{p}_{j\theta}(t)}{dt}=0.
\end{split}
\end{equation}
Performing the trace  gives us the following equations of motion for the expectation values of the operators
\begin{equation}
\label{eq:LangEq}
\begin{split}
&\frac{d\langle\hat{a}_{j}(t)\rangle}{dt}=-i(\Delta_{c_{j}}-i\beta_{j})\langle\hat{a}_{j}\rangle+i\xi_{j}\langle\hat{x}_{j}\hat{a}_{j}\rangle+\sqrt{\kappa_{ex}}(\varepsilon_{l_{j}}+\\
&\varepsilon_{p_{1}}e^{-i\eta_{j}t}\delta_{j1})-iJ_{j,j-1}\langle\hat{a}_{j-1}\rangle-iJ_{j,j+1}\langle\hat{a}_{j+1}\rangle,\\
&\frac{d^{2}\langle\hat{x}_{j}(t)\rangle}{dt^{2}}=-(\omega^{2}_{m_{j}}+\gamma_{m_{j}}\frac{d}{dt}   )\langle\hat{x}_{j}\rangle+\frac{\xi_{j}}{m_{j}}\langle\hat{a}^{\dagger}_{j}\hat{a}_{j}\rangle+\frac{\langle\hat{p}^{2}_{j\theta}\rangle}{m^{2}_{j}r^{3}_{j}},\\
&\frac{d\langle\hat{\theta}_{j}(t)\rangle}{dt}=\frac{\langle\hat{p}_{j\theta}\rangle}{m_{j}r^{2}_{j}},\\
&\frac{d\langle\hat{p}_{j\theta}(t)\rangle}{dt}=0.
\end{split}
\end{equation}
$\eta_{j}\equiv\omega_{p}-\omega_{l_{j}}$ and $\beta_{j}=1/2(\kappa_{ex}\delta_{j1}+\kappa^{(j)}_{in})$ is the net photon leakage rate from the $j$th cavity. Note that in the above set of equations we have phenomenologically added the optical and mechanical decay rates $\beta_{j}$ and $\gamma_{m_{j}}$, respectively. 
\subsection{Steady-state analysis}
Next, in order to proceed with the analytic solution we apply the so-called mean-field approximation \cite{agarwal2010OMIT}. To this end, we express all correlations as the product of the average value of operators. For instance we express $\langle \hat{x}_{j}\hat{a}_{j}\rangle= \langle \hat{x}_{j}\rangle \langle \hat{a}_{j}\rangle$ and $\langle \hat{a}^{\dagger}_{j}\hat{a}_{j}\rangle=\langle \hat{a}^{\dagger}_{j}\rangle \langle \hat{a}_{j}\rangle$. We then follow the standard procedure (see for instance \cite{jing2015optomechanically, *boyd2003nonlinear}) and decompose all operators as a sum of their steady-state value and small fluctuations around the steady-state value in the following form
\begin{equation}
\label{eq:SSansatz}
\begin{split}
& \langle\hat{a}_{j}\rangle\longrightarrow a_{j}+\delta a_{-j}e^{-i\eta_{j}t}+\delta a_{+j}e^{i\eta_{j}t},\\
& \langle\hat{x}_{j}\rangle\longrightarrow x_{j}+\delta x_{j}e^{-i\eta_{j}t}+\delta x^{\ast}_{j}e^{i\eta_{j}t}.
\end{split}
\end{equation}
By inserting Eq.~(\ref{eq:SSansatz}) into Eq.~(\ref{eq:LangEq}) we readily drive the steady-state values as
\begin{equation}
\label{eq:SSvalues}
\begin{split}
& a_{j}=\frac{(\sqrt{\kappa_{ex}}\varepsilon_{l_{j}}-iJ_{j,j-1}a_{j-1}-iJ_{j,j+1}a_{j+1})}{i\Delta_{c_{j}}-i\xi_{j}x_{j}+\beta_{j}},\\
& x_{j}=\frac{1}{m_{j}\omega^{2}_{m_{j}}}(\xi_{j}\mid a_{j}\mid^{2}+r_{j}\Omega^{2}_{j}).
\end{split}
\end{equation}
$|\Omega_{j}|=\frac{d\theta_{j}}{dt}$ is the magnitude of the spinning rate. Likewise, the fluctuating part of the expectation values of the operators can be worked out as
\begin{equation}
\label{eq:Flucvalues}
\begin{split}
& \delta a_{-j}(\beta_{j}+i\Delta_{c_{j}}-i\xi_{j}x_{j}-i\eta_{j})-i\xi_{j}a_{j}\delta x_{j}=\sqrt{\kappa_{ex}}\varepsilon_{p_{1}}\delta_{j1}-\\
&iJ_{j,j-1}\delta a_{j-1}-iJ_{j,j+1}\delta a_{j+1},\\
& \delta a^{\ast}_{+j}(\beta_{j}-i\Delta_{c_{j}}+i\xi_{j}x_{j}-i\eta_{j})+i\xi_{j}a^{\ast}_{j}\delta x_{j}=iJ_{j,j-1}\delta a^{\ast}_{j-1}+\\
&iJ_{j,j+1}\delta a^{\ast}_{j+1},\\
&(\omega^{2}_{m_{j}}-\eta_{j}-i\eta_{j}\gamma_{m_{j}})\delta x_{j}=\frac{\xi_{j}}{m_{j}}(a^{\ast}_{j}\delta a_{-j}+a_{j}\delta a_{+j}).
\end{split}
\end{equation}
In the derivation of Eq.~(\ref{eq:SSvalues}) and Eq.~(\ref{eq:Flucvalues}) we have adopted a perturbation approach where in all decompositions we have assumed the steady-state mean values to be much larger than the fluctuations i.e. $|a_{j}|>>|\delta a_{\pm j}|$ and $|x_{j}|>>\lbrace|\delta x_{j}|,|\delta x^{\ast}_{j}|\rbrace$.
\subsection{Probe field transmission rate}
The transmission rate $T$ of the probe field is defined as
\begin{equation}
\label{eq:ProbTrans}
T\equiv\mid t_{p}\mid^{2}=\frac{<\hat{a}^{\dagger}_{out}\hat{a}_{out}>}{<\hat{a}^{\dagger}_{in}\hat{a}_{in}>}.
\end{equation}
The probe field input ($\hat{a}_{in}$) and output ($\hat{a}_{out}$) operators are related through the standard Collett and Gardiner input-output relationship \cite{gardiner1985input}
\begin{equation}
\label{eq:InOut}
\hat{a}_{out}=\hat{a}_{in}-\sqrt{\kappa_{ex}}\delta a_{-1}=\varepsilon_{p_{1}}-\sqrt{\kappa_{ex}}\delta a_{-1}.
\end{equation}
Since probe is classical therefore we have replaced the input operator by the probe field amplitude $\varepsilon_{p_{1}}$ in Eq.~(\ref{eq:InOut}). Using Eq.~(\ref{eq:InOut}) in Eq.~(\ref{eq:ProbTrans}), the probe transmission rate can be expressed in terms of $\delta a_{-1}$ as
\begin{equation}
T=\Bigg| 1-\frac{\sqrt{\kappa_{ex}}}{\varepsilon_{p_{1}}}\delta a_{-1}\Bigg|^{2}.
\end{equation}
Hence, to find out the net transmission rate $T$, equation sets ~(\ref{eq:SSvalues}) and ~(\ref{eq:Flucvalues}) are simultaneously solved to obtain $\delta a_{-1}$. 
\section{Results and Discussion: the case of two spinning resonators}
The probe transmission derivation presented in the previous section is applicable to any number of resonators. In this section we concentrate on two-resonator problem in detail and study how different spinning directions can impact the probe transmission. To this end, we consider the following cases
\begin{itemize}
  \setlength{\itemsep}{0pt}
  \setlength{\parskip}{0pt}
  \setlength{\parsep}{0pt}
\item both resonators are non-spinning,
\item both resonators have clockwise spin,
\item $1^{st}$ has clockwise, $2^{nd}$ has counter-clockwise spin,
\item $1^{st}$ has counter-clockwise, $2^{nd}$ has clockwise spin,
\item only one of the two resonators is spinning,
\item and both resonators have counter-clockwise spin.
\end{itemize}
\subsection{Parameters}
We begin by mentioning the set of experimentally feasible parameters \cite{lu2017optomechanically, grudinin2010phonon, yang2015topological} considered in this work. Apart from spinning directions, we take all resonators to be identical with $m= 2$ng, $\omega_{m}=200$MHz, $\gamma_{m}=0.2$MHz, optical wavelength $= \lambda= 1.55\mu$m, refractive index $ =n =1.44$, speed of light in the optical medium $=v =3\times 10^{8}$/n, $\omega_{c}=193.5$ THz, quality factor of the optical resonator $=Q =3\times 10^{7}$, $\kappa_{ex}=\omega_{c}/Q$, $\kappa^{(j)}_{in} =\kappa_{ex}$, $P_{l} =10$W, $r =0.25$mm and $\xi=\omega_{c}/r$. The frequency of driving field for both resonators is also assumed to be the same. Here we take resonator-resonator coupling $J=\kappa_{ex}$. The strength of parameter $J$ can be controlled by altering the separation between the resonators. For instance, in 2012 Peng et al. \cite{peng2012photonic} have experimentally shown that by reducing the gap between two resonators to $\sim 5\mu$m a strong inter-resonator coupling can be established. This strong coupling leaves its signature on the transmission spectrum in the form of splitting of two resonances. For a relevant discussion on coupling between a whispering-gallery microdisk resonator with a tapered fiber see  \cite{bo2017controllable}.

\begin{figure*}
\centering
  \begin{tabular}{@{}cccc@{}}
    \includegraphics[width=2.5in, height=1.8in]{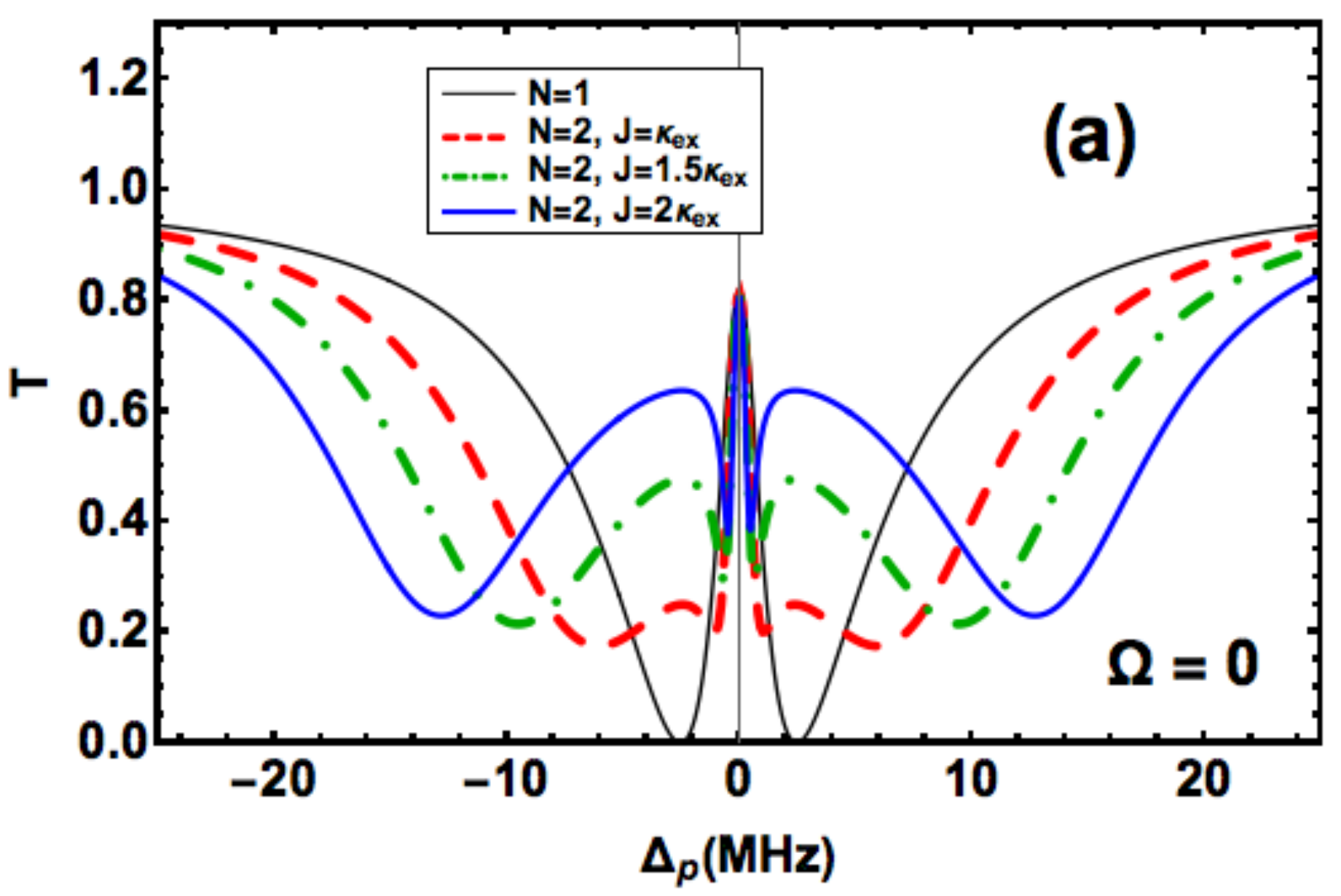} &
    \includegraphics[width=2.2in, height=1.82in]{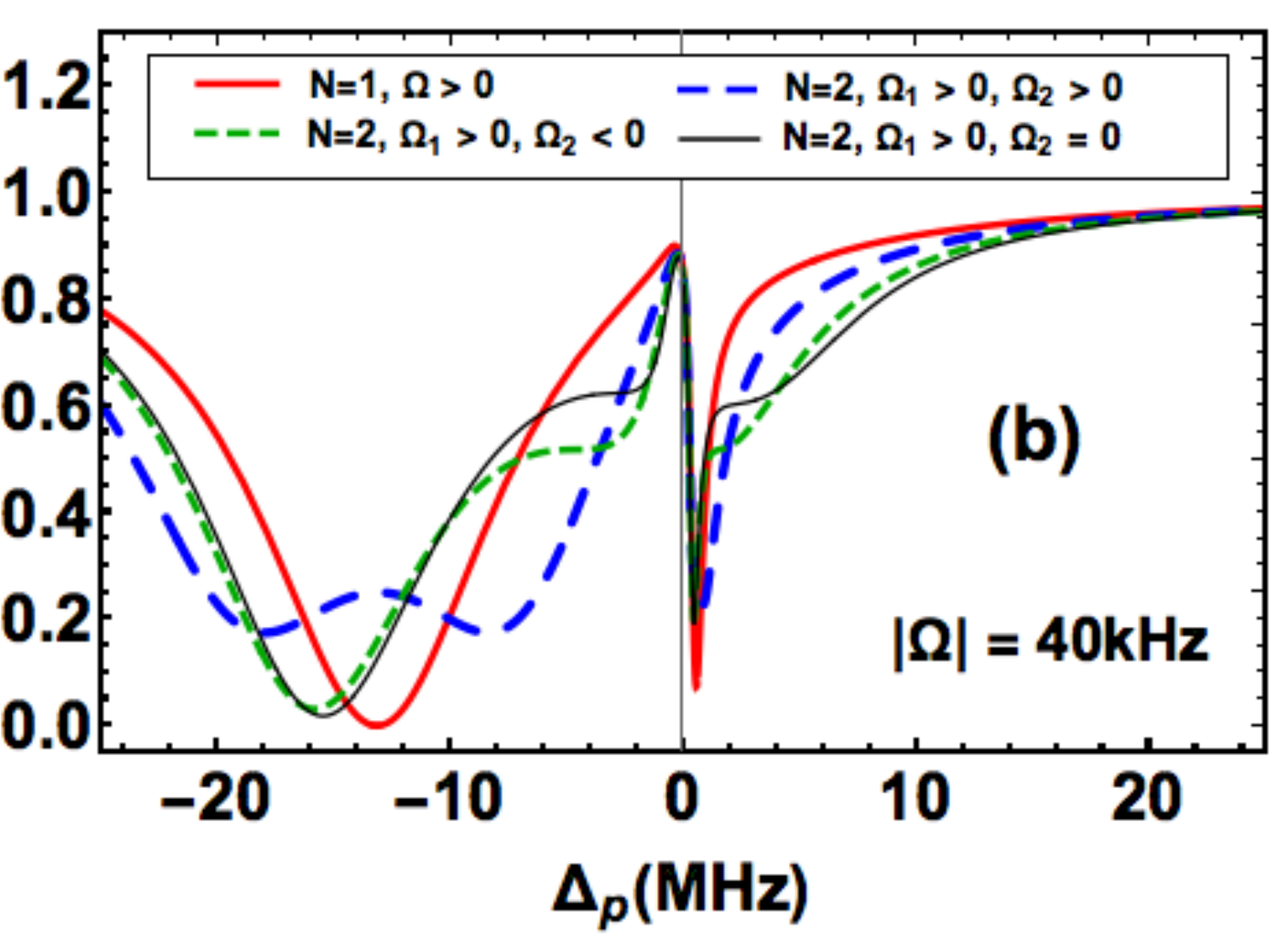} &
    \includegraphics[width=2.2in, height=1.821in]{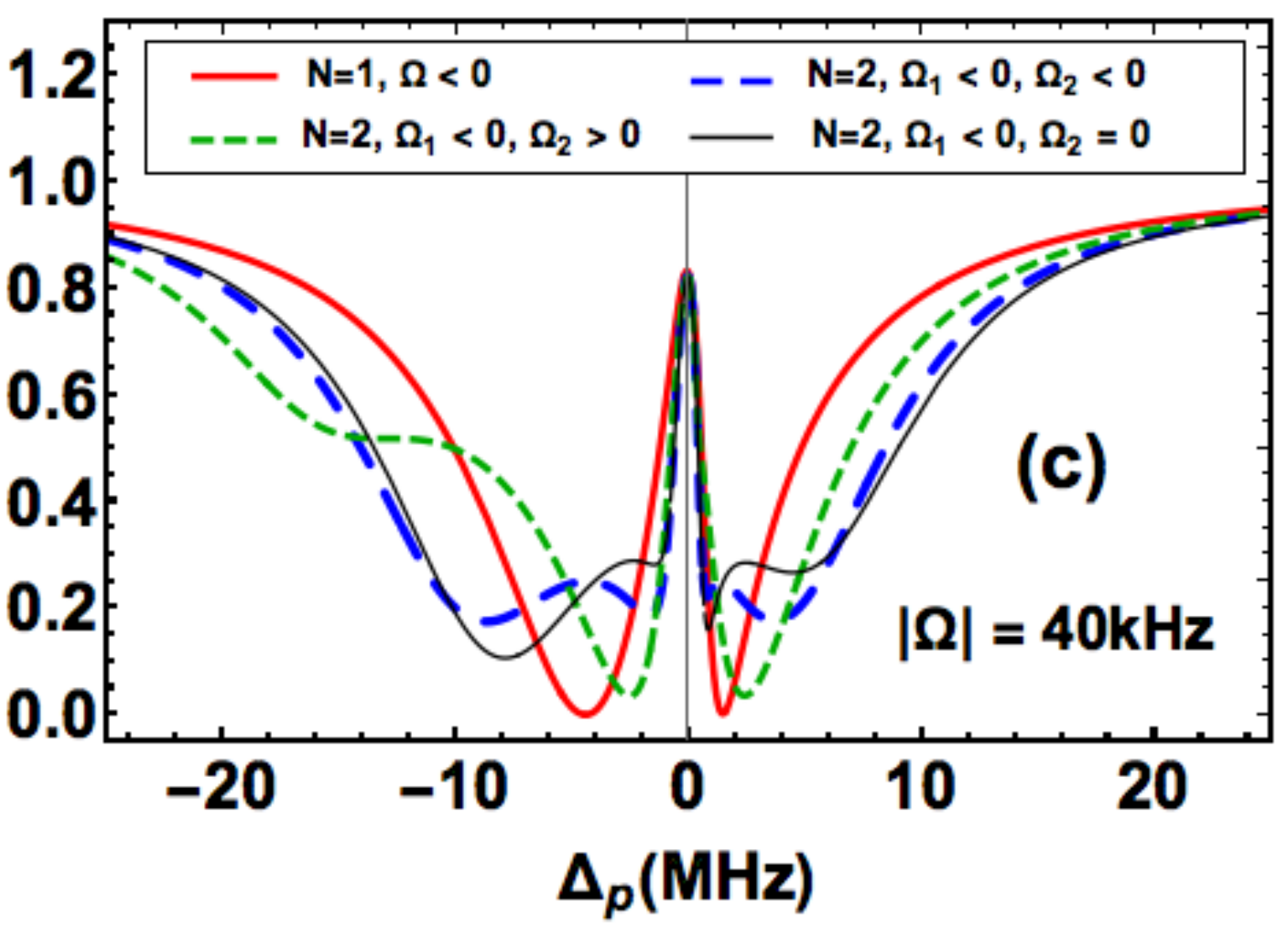} 
  \end{tabular}
\captionsetup{
  format=plain,
  margin=1em,
  singlelinecheck=false
}
 \caption{(Color online) Probe transmission rate as a function of detuning $\Delta_{p}$ for a single and a series coupled double resonator system. (a) Absence of spin  (b) Bottom resonator spinning in the clockwise direction (with rate $40$kHz) while upper resonator may or maynot be spinning in the same direction. We have also incorporated the scenario when the upper resoantor is not spinning. (c) Opposite situations to plot (b). In both (b) and (c) we have set $J/\kappa_{ex}=1$.}\label{Fig2}
\end{figure*}
\begin{figure*}
\centering
  \begin{tabular}{@{}cccc@{}}
   \hspace{-1mm}\includegraphics[width=2.5in, height=1.8in]{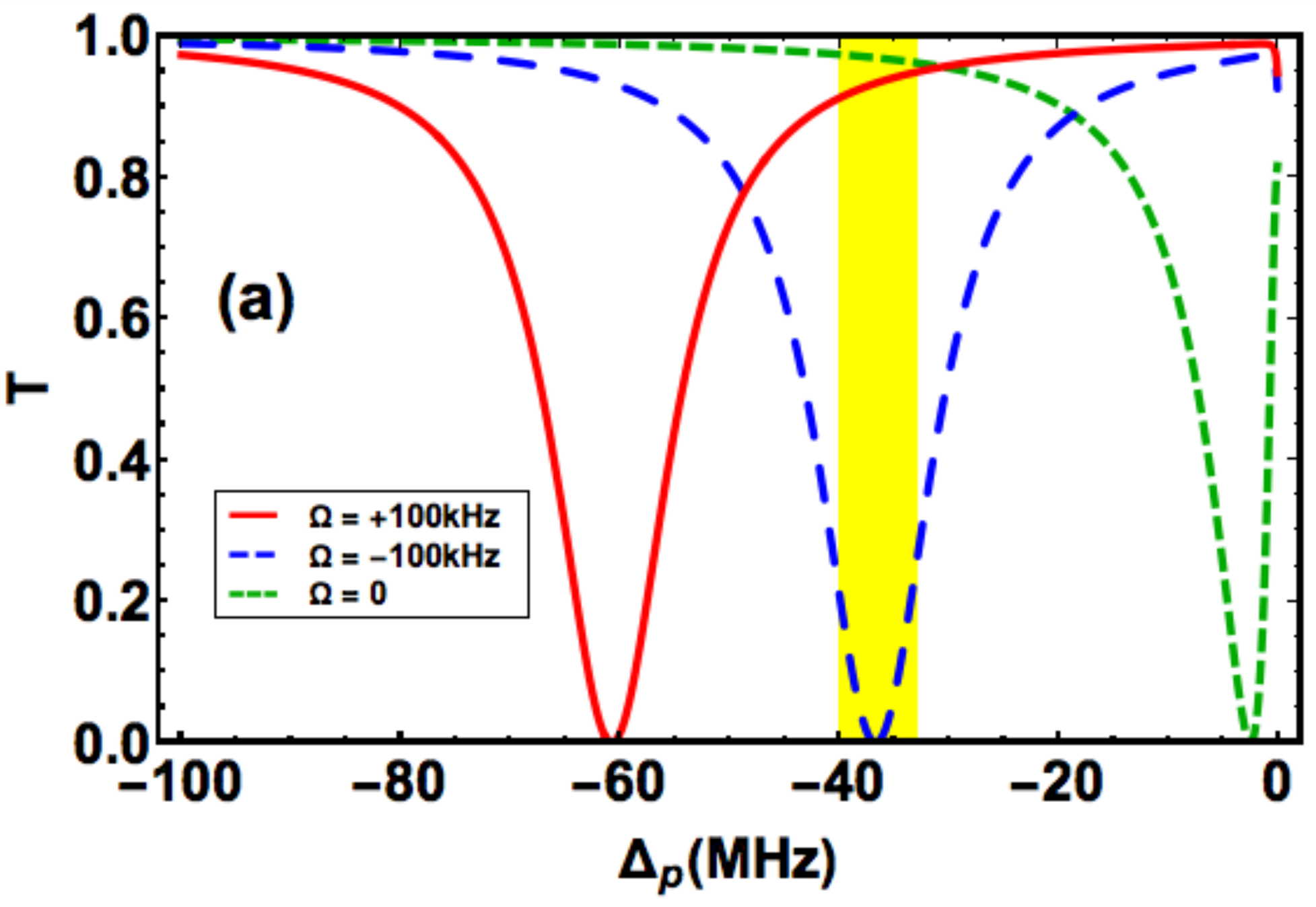} &
   \hspace{-1mm}\includegraphics[width=2.2in, height=1.8in]{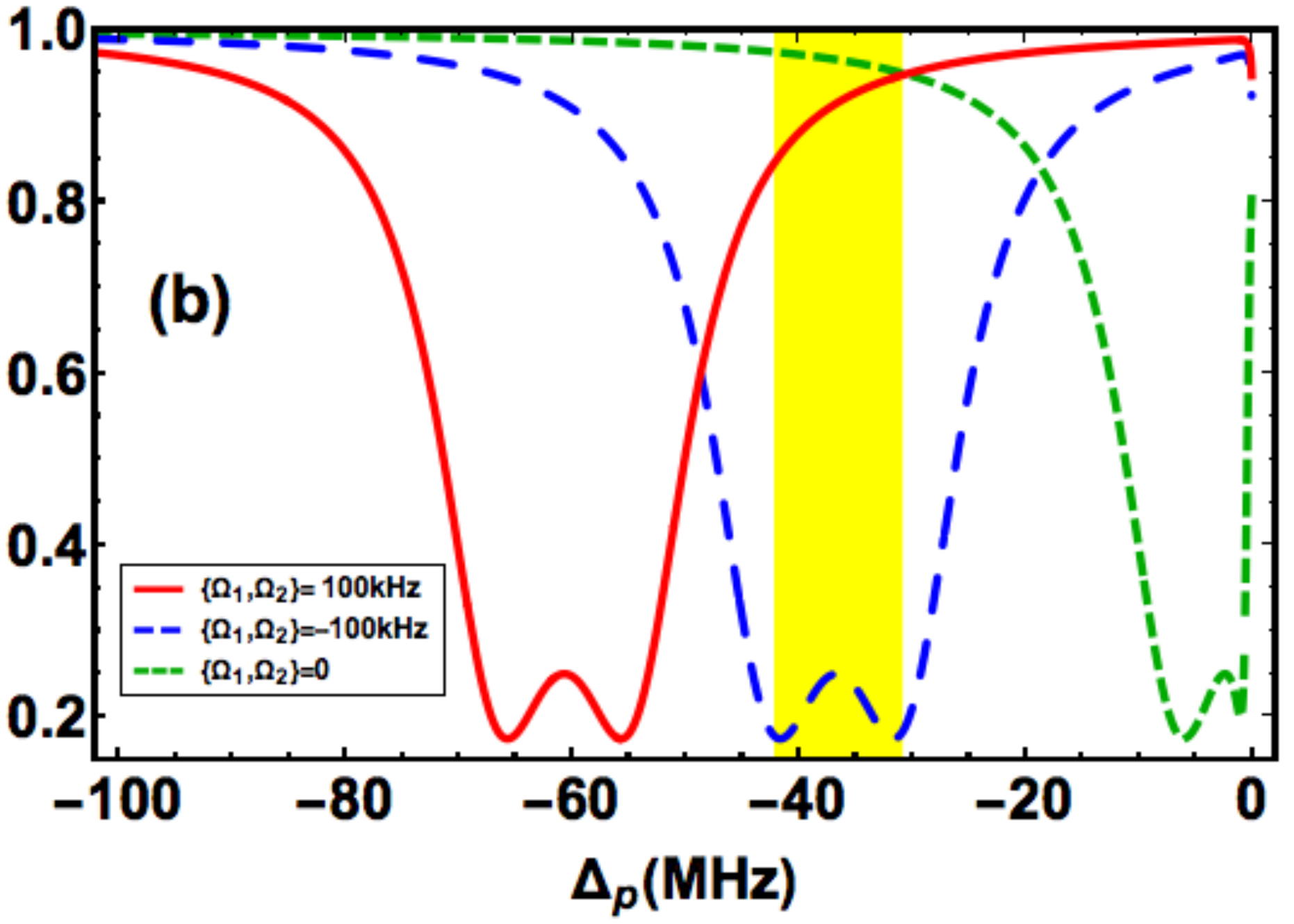} &
   \hspace{-1mm}\includegraphics[width=2.22in, height=1.791in]{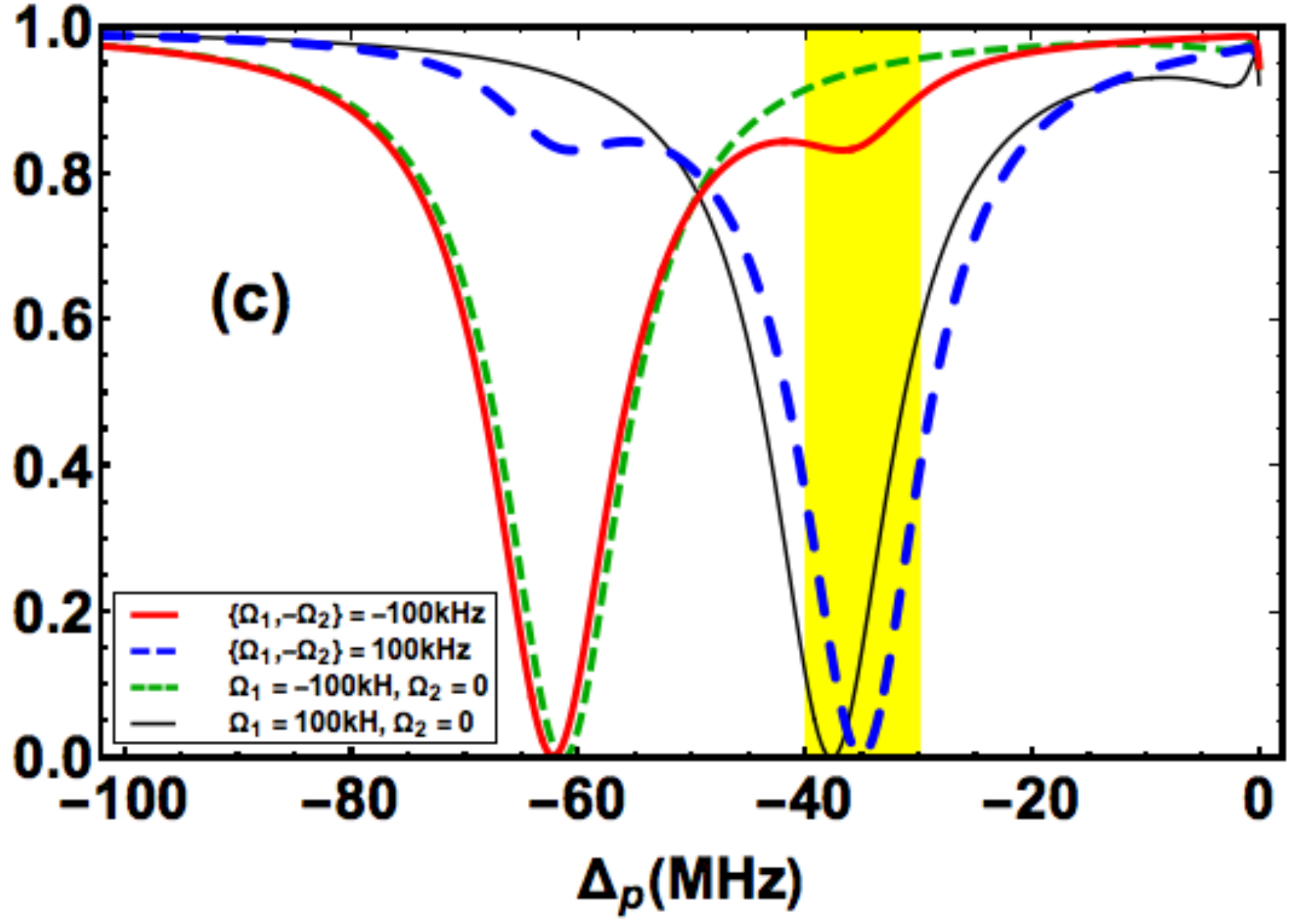} 
  \end{tabular}
\captionsetup{
  format=plain,
  margin=1em,
  justification=raggedright,
  singlelinecheck=false
}
 \caption{(Color online) Probe transmission rate for larger spinning rate ($|\Omega|=100$kHz). The yellow highlighted regions show the possibility of optical nonreciprocal transmission. (a) Single-resonator with different spinning directions (b) and (c) two coupled resonators with different combinations of spinning directions. }\label{Fig3}
\end{figure*}
\subsection{Probe transmission: non-reciprocal transport}
We start with Fig.~\ref{Fig2}(a) where we compare probe transmission of a single- and a double-resonator problem without the involvement of any spin (i.e. $\Omega_{1}=\Omega_{2}=0$). For a single resonator ($N=1$) case one finds
\begin{equation}
\delta a_{-1}=-\frac{ \sqrt{\kappa_{ex}}\varepsilon_{p_{1}}\lbrace i\xi^{2}\vert a\vert^{2}+m(\widetilde{\beta}_{-})\Gamma_{m}   \rbrace   }{  i\xi^{2}\vert a\vert^{2}(\widetilde{\beta}_{-})-(\widetilde{\beta}^{\ast}_{-})\lbrace i\xi^{2}\vert a\vert^{2}+m(\widetilde{\beta}_{-})\Gamma_{m}\rbrace   },
\end{equation}
where $\widetilde{\beta}_{-}+i\eta=(\beta-i\Delta_{c}+ix\xi)$, $\widetilde{\beta}^{\ast}_{-}+i\eta=(\beta+i\Delta_{c}-ix\xi)$ and $\Gamma_{m}=\omega^{2}_{m}-i\gamma_{m}\eta-\eta^{2}$. From the plot we notice the appearance of the standard OMIT transmission with peak residing at the resonance point ($\Delta_{p}\equiv\omega_{l}-\omega_{p}=0$).  As reported in \cite{weis2010optomechanically} the transparency window linewidth is given by $\gamma_{m}+\xi^{2}|a|^{2}/(m^{2}\omega^{2}_{m}\beta)$ which for our case takes the numerical value of $\sim2$MHz. For some related studies on non-reciprocal light propagation in non-spinning single optomechanical and optical resonators we direct reader to \cite{*hafezi2012optomechanically, *xia2014reversible, *ruesink2016nonreciprocity}. 
\\

For a double-resonator ($N=2$) setup without spin, we find that as we increase the resonator-resonator coupling strength ($J_{1,2}=J_{2,1}=J$) the whole transmission spectrum lifts upwards. However, the location of the OMIT peak stays unaffected. Additionally on each side of the OMIT peak two points of suppressed transmission are observed. For instance for $\Delta_{p}<0$ and $J=\kappa_{ex}$ (red dashed curve in Fig.~2(a)) two lowest $T$ values appear at $\Delta_{p}\sim -1$MHz and $\sim -7$MHz. Moreover, we find that the separation between the two lowest transmission points on positive or negative $\Delta_{p}$-axis can be enhanced by increasing the $J$ value. For instance for $J=2\kappa_{ex}$ ($\kappa_{ex}=6.45$ MHz) the separation between the lowest transmission points either on positive or negative $\Delta_{p}$-axis turns out to be almost $2\kappa_{ex}$ (see thick solid blue curve in Fig.~2(a)). Notice that the analytic expression of $\delta a_{-1}$ for two-resonators is mathematically involved and therefore not reported here.
\\

Next, in Fig.~2(b) and (c) we introduce the spin degree of freedom. Using Eq.~(\ref{eq:SSvalues}), for a single resonator case the steady-state values of $\langle \hat{a}\rangle$ and $ \langle\hat{x} \rangle$ follow
\begin{equation}
a=\frac{\sqrt{\kappa_{ex}}\varepsilon_{l}}{\beta+i\Delta_{c}-i\xi x}, \hspace{2mm} x=\frac{(\xi|a|^{2}+r\Omega^{2})}{m\omega^{2}_{m}}.
\end{equation}
Clearly when the resonator is capable of spinning ($\Omega\neq 0$), the value of $\xi x$ as well as $\Delta_{sag}$ modify which influence the full spectrum. In particular, we note that in both $\Omega>0$ and $\Omega<0$ situations the OMIT peaks are slightly red-shifted (move towards $\Delta_{p}<0$) as discussed in \citep{lu2017optomechanically}. For two spinning coupled resonators, the relevant steady-state values are modified due to resonator-resonator coupling
\begin{equation}
\begin{split}
&a_{1}=\frac{(\sqrt{\kappa_{ex}}\varepsilon_{l}-iJ_{1,2} a_{2})}{\beta_{1}+i\Delta_{c_{1}}-i\xi_{1} x_{1}},\hspace{2mm}a_{2}=\frac{(\sqrt{\kappa_{ex}}\varepsilon_{l}-iJ_{2,1} a_{1})}{\beta_{2}+i\Delta_{c_{2}}-i\xi_{2} x_{2}},\\
&x_{1}=\frac{(\xi_{1}|a_{1}|^{2}+r_{1}\Omega^{2}_{1})}{m_{1}\omega^{2}_{m_{1}}},\hspace{2mm} x_{2}=\frac{(\xi_{2}|a_{2}|^{2}+r_{2}\Omega^{2}_{2})}{m\omega^{2}_{m_{2}}}.
\end{split}
\end{equation}
As a result, we find that either the spin directions are the same or opposite, the transmission shows a different profile as compared to the single-resonator transmission. When both resonators are spinning in the same direction, for example clockwise, an important feature of the coupled resonators is the splitting of the spectrum due to resonant mode-coupling between the two resonators. As shown in Fig.~2(b), this leads to a unique ``W"-like transmission pattern with two symmetric dips for negative values of probe detuning $\Delta_p$ (see dashed blue curve in Fig.~2(b)). When the spinning directions are opposite, for example $\Omega_1>0$ and $\Omega_2<0$ (green-dashed curve in Fig.~2(b)), we observe asymmetric splitting of the spectrum due to off-resonant coupling between two cavity modes. When both spinning directions are inverted comparing to the previous cases, we observe similar features (Fig.~2(c)) but with different spectra due to the nonreciprocity of the spinning resonators. When we consider only one of the two resonators to be spinning (say $\Omega_{1}\lessgtr 0$ but $\Omega_{2}=0$) we find $T$ remains asymmetric. From Fig.~2(b) and (c), we notice that the one-spinning case can be quite different from the two-spinning case even if one of the two resonators in both situations have the same spinning direction (e.g., $\Omega_1$ has the same sign in blue dashed and thin black in Fig.~2(b)). These features extend down to higher spinning speeds as well (see Fig.~3). Overall these trends clearly demonstrate that the spin degree of freedom in a two spinning resonator problem gives us additional control and thus can be used for the probe transmission alteration.

As discussed in \citep{lu2017optomechanically} for a single resonator problem with $|\Omega|=100$kHz and $\xi x=48.47$ MHz (Fig.~3(a)), the spinning direction can be used to tune the position of the OMIT peak which for $\Omega=0$ case resides at $\Delta_{p}=0$. Moreover, at the same $\Delta_{p}$ value we can achieve $T> 0.9$ (pass) and $T\sim 0$ (block) by adjusting the spinning directions in the clockwise and counter-clockwise directions, respectively for a left incoming probe field. We find a second coupled resonator can introduce new features in $T$ versus $\Delta_{p}$ profile at higher spinning rates. For example, when both resonators spin in the same direction (Fig.~3(b)),  a ``W" like transmission appears which can be controlled by changing the sign of the spinning directions and value of $J$. In this situation with $\xi x=48.47$ MHz, in the highlighted region when both resonators spin in the clockwise direction $T$ takes $90$\% maximum value (solid red curve in Fig.~3(b)). But when the spinning direction is changed to counter clockwise direction the same probe field incoming from left suffers minimum transmission (blue dashed curve). There  we also obtain two minima of $T$ where transmission is suppressed to less than $20$\%. Additionally, in between two minima we observe a local maximum with $T\gtrsim25$\%. When the second resonator is non-spinning $T$ becomes slightly red shifted. These features are shown more clearly in Fig.~4 (compare thin dashed and thin solid red curves).

\begin{figure}
\centering
   \includegraphics[width=3.3in, height=2.3in]{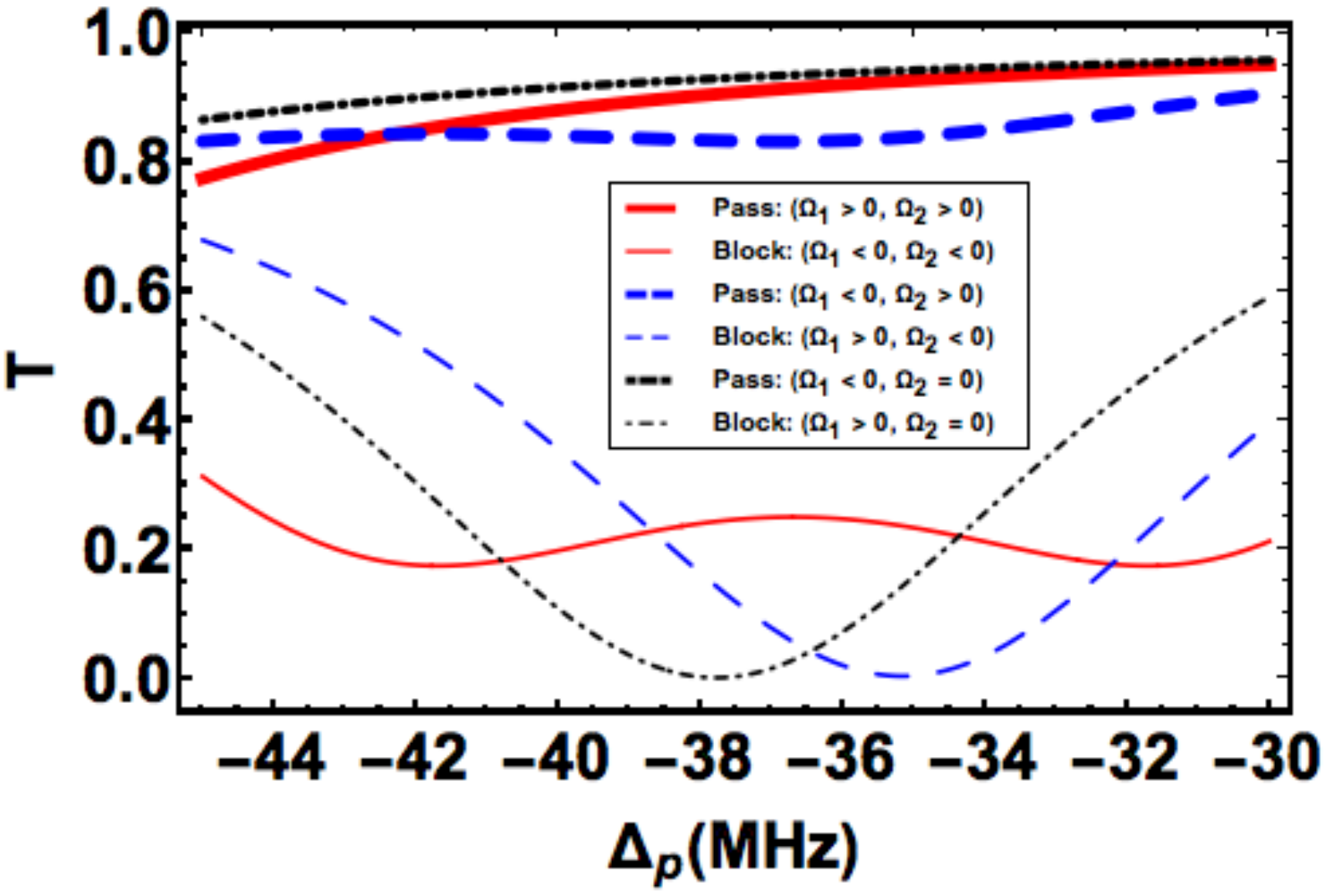} 
\captionsetup{
  format=plain,
  margin=1em,
  justification=raggedright,
  singlelinecheck=false
}
 \caption{(Color online) Magnified version of highlighted regions of Fig.~3(b) and (c) plotted on the same scale.}\label{Fig4}
\end{figure}

Finally, in Fig.~3(c) we address the scenario in which both resonators spin in different directions. Here (as shown in the yellow highlighted region of Fig.~3(c) and in Fig.~(4)), the minimum transmission achieves zero value when the bottom resonator spin in the clockwise direction. Whereas, if the spin of the resonators is reverted then the transmission becomes maximum with  $T>80$\% around $\Delta_{p}=35$MHz in all cases. As an example, we notice Fig.~3(b), (c) and Fig.~4 clearly show that, if $\Omega_{1}<0, \Omega_{2}<0$ specifies the case when pump and probe are applied from the right direction then flipping the applied probe direction to the left and choosing appropriate $\Delta_{p}$ value we can obtain non-reciprocal probe transmission.

\subsection{Transmission rate enhancement factor (E.F.)}
\begin{figure}[h]
\centering
   \includegraphics[width=3.5in, height=2.25in]{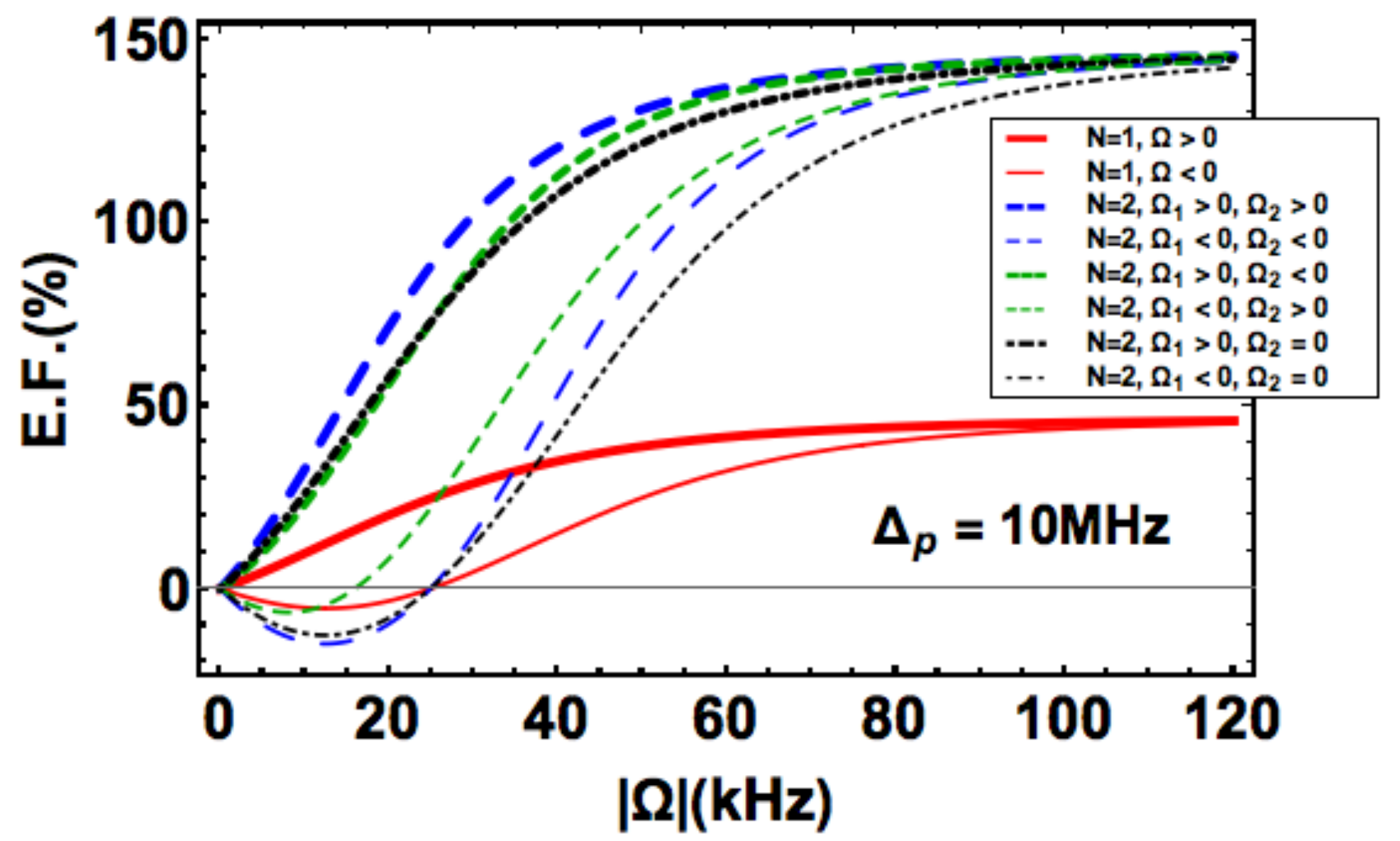} 
\captionsetup{
  format=plain,
  margin=1em,
  justification=raggedright,
  singlelinecheck=false
}
 \caption{(Color online) Off-resonance transmission enhancement factor as a function of spinning rate for various spinning directions in single- and double-resonator setups. Note that for $N=2$ we have assumed $\vert\Omega_{1}\vert=\vert\Omega_{2}\vert=\vert\Omega\vert$. Rest of the parameters are same as used in Fig.~2.}\label{Fig5}
\end{figure}
In order to quantify how much transmission is augmented when two resonators take different spinning directions at a fixed probe detuning $\Delta_{p}$ value, we define the transmission rate enhancement factor (E.F.) as
\begin{equation}
E.F.=\frac{T(\Omega_{1}\neq 0, \Omega_{2}\neq 0)}{T(\Omega_{1}=0,\Omega_{2}=0)}\Bigg\vert_{\Delta_{p}}-1.
\end{equation}
For a single spinning resonator the expression of the E.F. simplifies to
\begin{equation}
E.F.=\frac{T(\Omega\neq 0)}{T(\Omega=0)}\Bigg\vert_{\Delta_{p}}-1.
\end{equation}
From the inspection of Fig.~2 we notice that under all spinning directions $T$ takes almost the same value at the OMIT point ($\Delta_{p}=0$). Therefore, we choose an off-resonant value of $\Delta_{p}=10MHz$ to clearly show considerable change in the maximum value of the probe transmission through the E.F. It is worthwhile to mention that the E.F. is sensitive to the choice of the $\Delta_{p}$ value such that for a different $\Delta_{p}$ we can obtain results completely opposite to what are presented below. The goal in choosing $\Delta_{p}=10$MHz here is to highlight a case where transmission is enhanced.

For a single resonator case, we notice as we increase $|\Omega|$ value for $\Omega>0$ scenario (thick red solid curve), E.F. shows growth such that for $|\Omega|\geq 80$kHz E.F. takes $\sim 45$\% value. On the contrary, the corresponding $\Omega<0$ scenario (thin red solid curve) shows a declining trend (negative E.F. value) up to $|\Omega|\approx 25$kHz. Crossing this point, E.F. becomes positive and joins the E.F. asymptotic value achieved in the $\Omega>0$ case.

For the corresponding $N=2$ case, the maximum E.F. as compared to single resonator situation is improved by a factor of 3 at $100$kHz. However, the counter-clockwise spin of the bottom resonator ($\Omega_{1}<0$) degrade higher transmission such that E.F. remains negative up to $|\Omega|\approx 15$kHz for $\Omega_{2}>0$ and $25$kHz for $\Omega_{2}<0$ and $\Omega_{2}=0$ cases, respectively. When $|\Omega|$ values is further raised the E.F. increases and finally (independent of spinning directions) reaches its maximum of $\sim150$\%.

\subsection{Group delay: slow and fast light control}
In order to characterize slow light and fast light propagation \cite{safavi2011elect}, we use probe field group delay $\tau_{g}$ as a quantifying parameter. It is defined as
\begin{equation}
\tau_{g}=\frac{d{\rm arg}(t_{p})}{d\Delta_{p}}
\end{equation}
Consequently one can also define the group delay enhancement factor $G.D.$ for two-resonators as
\begin{equation}
G.D.=\frac{\tau_{g}(\Omega_{1}\neq 0,\Omega_{2}\neq 0)}{\tau_{g}(\Omega_{1}=0,\Omega_{2}=0)}\Bigg\vert_{\Delta_{p}}-1.
\end{equation}
\begin{figure}
\centering
   \includegraphics[width=3.4in, height=2.3in]{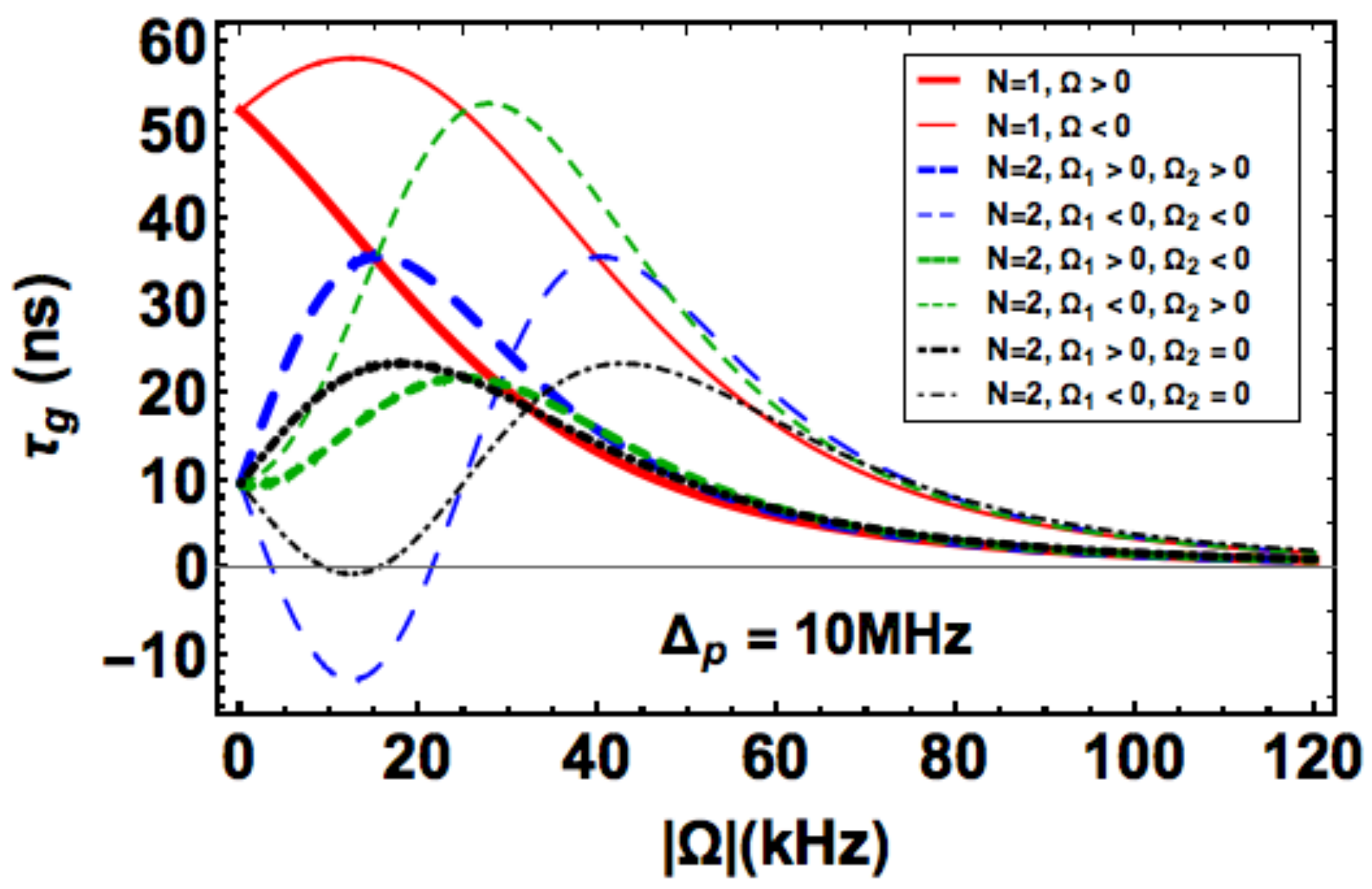} 
\captionsetup{
  format=plain,
  margin=1em,
  justification=raggedright,
  singlelinecheck=false
}
 \caption{(Color online) Group delay of probe light as a function of spinning rate magnitude for various spin direction options. Parameters are the same as used in Fig.~2.}\label{Fig6}
\end{figure}

Here we focus on the group delay itself and plot $\tau_{g}$ in Fig.~6 as we vary the spinning rate $|\Omega|$ with $\Delta_{p}=10$MHz.
In all cases we notice that the behavior of $\tau_{g}$ is sensitive to $\Delta_{p}$ and $|\Omega|$ values. For a single counter-clockwise spinning resonator (thin solid red curve), $\tau_{g}$ gradually increases and reaches a maximum around $15$kHz and then decays to almost zero value as we increase the spinning rate to $120$kHz. On the other hand, for double resonators with $\Omega_{1}<0$ and $\Omega_{2}<0$ situation can lead to negative $\tau_{g}$ values for $|\Omega|\lesssim25$kHz. In another case when $\Omega_{1}>0$ and $\Omega_{2}=0$ group delay becomes slightly negative around $10$kHz and then shows sinusoidal-like behavior. All of these cases show the possibility of spinning related slow and fast light propagation with coupled resonators. 

For a two-resonator problem, one can compare Fig.~(\ref{Fig5}) and Fig.~(\ref{Fig6}) and point out different $|\Omega|$ values where in addition to high transmission slow or fast light can be achieved. For instance, when $|\Omega|=20$kHz and $\Omega_{1}>0,\Omega_{2}>0$ (thick blue dashed curves), one can achieve both enhanced transmission ($E.F.\sim 60$\%) as well as considerable slow light ($\tau_{g}\sim 35$ns). Similarly if we choose $\Omega_{1}<0,\Omega_{2}<0$ (thin blue dashed curve) at $|\Omega|=50$kHz probe light group velocity will be lowered in addition to the enhanced transmission. For $\Omega_{2}=0$, $\Omega_{1}<0$ case (black dotted dashed curve) at $40$kHz we find $E.F.\sim 40\%$ with significant time delay of $\tau_{g}\sim 20$ns. This shows that the availability of a second spinning resonator can be utilized to prolong/store output probe light transmission. This feature can then be used in quantum memory component of the quantum communication protocols.

\section{Summary}
To recapitulate, we investigated the probe transmission properties in series coupled spinning optomechanical resonators. For $\Omega=0$ case, single resonator leads to OMIT. The presence of a second resonator lifts the transmission upwards as well as enhances the separation between lowest value of T as resonator-resonator coupling ($J$) value is elevated. When $\Omega\neq 0$ was considered we found the value of $\xi x$ changes for both $N=1$ and $N=2$ cases. This, for instance, led to spin direction-dependent OMIT peak shift for $N=1$ case and the emergence of a ``W" like pattern for $\Delta_{p}<0$ values for $N=2$ scenario. Further, the transmission E.F. showed the double resonator raised the maximum transmission by a factor of $3$ as compared to $N=1$ situation for $\Delta_{p}=10$MHz at $|\Omega|=100$kHz. The case in which we set $\Omega_{2}=0$ and $\Omega_{1}\neq0$ the transmission and $E.F.$ retain an asymmetric profile. Therein, we find that by switching on-and-off the spinning of the second resonator provides an additional tunability in the transmission profile.

When we selected a higher spinning rate for both resonators ($|\Omega_{1}|=|\Omega_{2}|=100$kHz) and focused around $\Delta_{p}=40$MHz, we noticed that the same and opposite rotary directions produced oscillations in the curves corresponding to the minimum and maximum transmission regions. This indicated the possibility of nonreciprocal probe light transmission by altering the probe launching direction. Finally, we studied slow and fast light propagation characterized by the group delay parameter ($\tau_{g}$). We focused on different choices of parameters where both high transmissions, as well as slow light, can be achieved. For $N=2$ case, one such example was when both resonators spin in the clockwise direction at $|\Omega|=20$kHz. In this case, we found a high transmission ($E.F.\sim 60$\%) and slow light ($\tau_{g}\sim 35$ns) can be simultaneously realized. \\
Here we would like to mention that in this work we consider all resonators to support optomechanical interactions but there are other interesting variations possible in this setup. For instance one can also study the case with coupled resonators, where one resonator is purely optical spinning resonator, while the other is a non-spinning optomechanical cavity supporting a mechanical mode. With such a system, one can for instance study the nonreciprocal amplification of phonons \cite{Jing_2018}. As a possible future extension, we envisage that this work can be extended to higher dimensions (say two or three dimensions), where resonators' spin can be utilized as an additional degree of freedom for the tunable optical transmission in coupled spinning optomechanical lattices.

\acknowledgments
I. M. M. would like to thank Miami University College of Arts and Science Start-Up funding. H. J. is supported by the National Natural Science Foundation of China (NSFC, 11474087 and 11774086).

\bibliographystyle{ieeetr}
\bibliography{paper.bib}

\begin{thebibliography}{10}
\bibitem{Jing_2018}
We came to know about such a setup and its applications through private
  communication with Prof. Hui Jing group.
\end{thebibliography}
\end{document}